\documentstyle[eqsecnum,preprint,aps]{revtex}

\begin{document}
\draft
\preprint{Th-October,1997}
\title{A Canonical Realization of the BMS Algebra}
\author{G.Longhi and M.Materassi}
\address{Department of Physics, University of Firenze}
\date{\today}
\maketitle

\begin{abstract}
  
A canonical realization of the BMS (Bondi-Metzner-Sachs) algebra is given 
on the phase space of the classical real Klein-Gordon field . 

By assuming the finiteness of the generators of the Poincar\'e group, it 
is shown that a countable set of conserved quantities exists 
(supertranslations); this set transforms under a particular Lorentz 
representation, which is uniquely determined by the requirement of 
having an invariant four-dimensional subspace, which corresponds to the 
Poincar\'e translations. This Lorentz representation is infinite-dimensional, 
non unitary, reducible and indecomposable. Its representation space is 
studied in some detail. It determines the structure constants of the 
infinite-dimensional canonical algebra of the Poincar\'e generators 
together with the infinite set of the new conserved quantities. 

It is shown that this algebra is isomorphic with that of the BMS group.

\end{abstract}
\bigskip


\narrowtext

\section{Introduction}
 
The purpose of this paper is to give a canonical realization of the algebra 
of the Bondi-Metzner-Sachs (BMS) group \cite{BMS} on the phase space of 
the real classical Klein-Gordon field. 

The BMS group arises in general relativity as the asymptotic symmetry at 
null infinity of a space-time describing an isolated (radiating) gravitational 
source. Such space-time is expected to become flat at null infinity 
\cite{FLAT}, and to exhibit the Poincar\'e group as the group of asymptotic 
symmetry.

While an asymptotic symmetry exists, its group is not the Poincar\'e 
group. It is a larger group containing the Poincar\'e group as a subgroup, 
and is a semidirect product of the homogeneous Lorentz group with an 
infinite dimensional abelian group. Among the infinitely many generators 
of this abelian group there are the generators of translations; the 
remaining are the so-called supertranslations. 

The idea of asymptotic flatness was formalized by Penrose by a process of 
conformal compactification \cite{PEN}; by means of this process a boundary 
(scri) was added to space-time, consisting of end points of null geodesics. 
In this way the notion of flatness acquired an intrinsic meaning. 

The asymptotic symmetry at null infinity has been studied by several authors. 
Apart from the classic papers \cite{BMS} we also quote some review articles 
\cite{BMSREVIEW}. Here we only mention some points discussed in the 
literature on this argument. 

One problem with the BMS group is due to the presence of infinitely many 
copies of the homogeneous Lorentz group, obtained from one another by 
conjugation with a supertranslation. This happens even in the case of the 
Poincar\'e group, but the spin Casimir of the Poincar\'e group is not 
invariant under such conjugation \cite{SACHS}. This causes difficulties 
in the definition of a unique Poincar\'e group, and, by consequence, in the 
definition of the angular momentum of the system \cite{WIN} (see however 
\cite{MOR}).

The unitary and irreducible representations of the BMS group have been 
studied in \cite{McC} to give a possible interpretation of this group 
as a fundamental group of elementary particles, a line of research no 
longer pursued.

There is a situation analogous to that of null infinity in the case of 
spatial infinity, where a similar (bigger) asymptotic symmetry arises 
\cite{SPI}.

More recently the classical approach to null infinity was criticized 
\cite{POLY}, and it was shown that a more general asymptotic expansion for 
the metric should be used (polyhomogeneous expansion). Nevertheless, the 
group of asymptotic symmetry at null infinity is still the BMS group.

In view of the role played by the BMS group in general relativity it seems 
of interest to show that its algebra can be canonically realized in terms of 
conserved quantities of a real classical Klein-Gordon field. While this 
field has been  chosen for its simplicity we expect that a similar analysis 
could be done for any other classical field. In particular, we will give 
explicitly the underlying infinite-dimensional representation 
of the Lorentz algebra, which determines the structure constants of the BMS 
algebra. Indeed we will show that, besides the total energy-momentum 
$P^{\mu}$ of the Klein-Gordon field, it is possible to define a countable 
set of other conserved quantities, with vanishing Poisson brackets with 
$P^{\mu}$ and among themselves. The requirement of the finiteness of 
$P^{\mu}$ implies a bound on the asymptotic behaviour of the Fourier 
coefficients of the field, when infinity is approached in the momentum space. 
This asymptotic behaviour is the same as required for the finiteness of 
this new set of conserved quantities. 

More precisely, the Fourier coefficients of the field are defined on 
the mass hyperboloid, which is a three-dimensional Riemannian 
manifold with constant negative curvature. On this manifold the 
Laplace-Beltrami operator is naturally defined. It is the only 
non-vanishing Casimir operator of the Lorentz group on the space of 
functions on this manifold. It is by the study of the eigenfunctions of 
this operator that we discover the required representation of the Lorentz 
group, which, as we will see, is an infinite-dimensional, non-unitary and 
reducible representation.

The four functions $k^{\mu}$ on the mass hyperboloid are eigenfunctions 
of the Laplace-Beltrami operator. In addition there exists a countable set 
of eigenfunctions, belonging to the same eigenvalue, which are, like $k^{\mu}$, 
not square integrable, with respect to the Lorentz invariant measure. This 
set, together with $k^{\mu}$, transforms according to the representation, 
and, what is crucial, {\sl{all}} these functions have the {\sl{same}} 
asymptotic behaviour. The functions $k^{\mu}$ span   an invariant 
four-dimensional subspace of this representation, but the orthogonal 
complement of this subspace is not invariant. So this representation 
is reducible but {\sl{not decomposable}}. 

The structure of this representation function space is the same as that of 
the translations and supertranslations of the BMS group. We show that the 
Poisson algebra of the conserved quantities built from these functions and 
the generators of the Lorentz group is the same as that of the BMS group. 
So we call these conserved quantities the generators of supertranslations. 

The action of the Lorentz generators on the supertranslations defines a set 
of structure constants of the BMS algebra, which are the matrix elements 
of the infinite-dimensional representation of the Lorentz group. 
The action of the supertranslations on the field is non-local. Nevertheless 
it defines, as in the case of $P^{\mu}$, a canonical transformation on 
the field. 

In Section II we recall some definitions about the Klein-Gordon field 
and define the Laplace-Beltrami operator on the mass hyperboloid. 
The eigenfunctions of this operator are studied in Section III and, 
in Section IV, the infinite representation of the Lorentz 
algebra is studied in some detail. In Section V the Poisson 
algebra of the generators of supertranslations is given and it is shown 
that it is isomorphic with the algebra of the BMS group.

The definitions and notations for the Klein-Gordon field are given in 
Appendix A. In Appendix B we discuss the eigenfunction problem for the 
Laplace-Beltrami operator on the mass hyperboloid. Moreover, we study 
in some detail the representation of the Lorentz group, which is the 
basis for the definition of the supertranslation generators.

\section{The Klein-Gordon field and the Laplace-Beltrami operator}

The notations and definitions for the real Klein-Gordon field 
$\Phi({\vec x},t)$ are given in Appendix A. For each $t \in R$ 
the field $\Phi({\vec x},t)$ 
will be supposed to belong, together with its spatial and temporal 
derivatives, to $L^2(R^3)$, that is 

\begin{equation}
(\Phi(\cdot,t),{\dot\Phi}(\cdot,t)) \in H^1(R^3)\times L^2(R^3),
\nonumber\end{equation}

\noindent where ${\overline{\Phi}} = {\partial \Phi\over \partial t}$ and 
where $H^1(R^3)$ is the $W^1_2(R^3)$ Sobolev space \cite{HAM}. 
This implies the existence of the total energy-momentum

\begin{equation}
P^{\mu} = \int d^3x \left[\dot\Phi({\vec x},t) \partial^{\mu}\Phi({\vec x},t) - 
{1\over 2}\eta^{\mu 0}
(\partial_{\nu}\Phi({\vec x},t)\partial^{\nu}\Phi({\vec x},t) - 
m^2\Phi^2({\vec x},t))\right],\label{2.100}\end{equation}

\noindent or, in terms of the Fourier coefficients of the field $\Phi$ 

\begin{equation}
P^{\mu} = \int\tilde{dk} k^{\mu} {\overline{a}}(\vec k) a(\vec k),
\label{2.110}\end{equation}

\noindent where the measure $\tilde{dk}$ is defined in Appendix A and 
$k^0 = \omega(k)$. The energy-momentum $P^{\mu}$ is time-like and future 
oriented.

Apart the existence of a well defined total momentum $P^{\mu}$, we require 
the existence of the generators of the Lorentz group. This implies a more 
stringent condition on $\Phi$, or on $a(\vec k)$. We require the 
additional condition: $\vec\nabla a(\vec k) \in L^2(R^3)$. 
From Eq.(\ref{2.110}), with $\mu = 0$, we have that $a(\vec k) \in L^2(R^3)$; 
with this additional condition we have that $a(\vec k)$ is continuous and 
that has a vanishing limit when $\vert\vec k\vert \rightarrow \infty$ 
\cite{Rich}. 

From Eq.(\ref{2.110}) we see that the asymptotic behaviour when 
$\vert\vec k\vert \rightarrow \infty$ of $a(\vec k)$ must be

\begin{equation}
\vert a(\vec k)\vert \simeq \vert\vec k\vert^{-{3\over 2}-\epsilon}
\label{2.120}\end{equation}

\noindent for any $\epsilon > 0$.

This means that any integral like (\ref{2.110}), with some regular function 
of $\vec k$ in place of $k^{\mu}$, and with the same degree of growth at 
infinite, will exist, as will be explicitly seen in Section VI. 

The scalar field $\Phi(x)$, where $x \equiv ({\vec x}, t)$, transforms under 
a Poincar\'e transformation $U(\Lambda,\alpha)$ as

\begin{equation}
(U(\Lambda,\alpha)\Phi)(x) = \Phi(\Lambda^{-1}(x-\alpha)).
\label{2.130}\end{equation}

This induces the transformation on $a(\vec k)$

\begin{equation}
(U(\Lambda,\alpha)a)(\vec k) = a(\Lambda^{-1}\vec k) e^{i(k\cdot \alpha)}.
\label{2.140}\end{equation}

In this last equation the notation $\Lambda^{-1}\vec k$ has the meaning 
$\Lambda^i_{\cdot \nu} k^{\nu}, i=1,2,3$, where 
$k^{\circ} = \sqrt{m^2 + {\vec k}^2}$. 

The canonical action of the Poincar\'e generators on $a(\vec k)$ is given by 
(see reference \cite{HAM})
\begin{equation}
\{ P^{\mu},\ a(\vec k)\} = i k^{\mu} a(\vec k),\label{2.150}
\end{equation}

\begin{equation}
\{ M^{\prime\mu\nu},\ a(\vec k)\} = D_{\mu\nu} a(\vec k),\ 
\{ M^{\prime\mu\nu},\ {\overline{a}}(\vec k)\} = D_{\mu\nu} {\overline{a}}(\vec k),
\label{2.160}\end{equation}

\noindent where $M^{\prime\mu\nu}$ is defined in Eq.(\ref{A.310}) and 

\begin{equation}
D_{\mu\nu} = (\eta_{\mu}^i k_{\nu} - 
\eta_{\nu}^i k_{\mu}){\partial\over \partial k^i}.
\label{2.170}\end{equation}

The differential operators $D_{\mu\nu}$ satisfy the algebra

\begin{equation}
[D_{\mu\nu},\ D_{\rho\lambda}] = 
\eta_{\mu\rho}D_{\nu\lambda} + \eta_{\nu\lambda}D_{\mu\rho} -
\eta_{\mu\lambda}D_{\nu\rho} - \eta_{\nu\rho}D_{\mu\lambda}.
\label{2.180}\end{equation}

We can now work with the field $a(\vec k)$ and its complex conjugate 
$\overline{a}(\vec k)$, which are defined on the mass hyperboloid. This 
is a Riemannian manifold which will be called $H^1_3$, following 
the notations of \cite{RLN}, where such manifolds are studied. 

Indeed, if we define $H^1_3$ as the inclusion f of the submanifold 
$q^2 - m^2 = 0$ with $q^0 > 0$, in the Minkowski space $M^4$ (with 
coordinates $q^{\mu}$), with f defined by

\begin{equation}
f: \{k^i\}\rightarrow \{q^{\mu}\},\quad {\rm with}\quad q^{\circ} = \sqrt{m^2 
+ \vec{k}^2},\quad {\rm and}\quad q^i = k^i,\quad (i = 1,2,3),\label{2.200}
\end{equation}

\noindent we get for the induced metric $\hat\eta$ 

\begin{equation}
\hat\eta = f^* \eta,\label{2.190}
\end{equation}

\noindent the following expression

\begin{equation}
{\hat\eta} = {\hat\eta}_{ij} dk^i dk^j,\label{2.210}\end{equation}

\noindent with 

\begin{equation}
{\hat\eta}_{ij} = {1\over \omega^2(k)} k^i k^j - \delta_{ij},\quad 
(i, j = 1,2,3).\label{2.220}
\end{equation}

The inverse of the matrix $\{ \hat\eta_{ij}\}$ is 

\begin{equation}
{\hat\eta}^{ij} = - (\delta^{ij} + {k^i k^j\over m^2}),\label{2.230}
\end{equation}

\noindent its determinant is

\begin{equation}
\mid{\hat\eta}\mid = det\{\hat\eta_{ij}\} = 
- {m^2\over \omega^2(k)},\label{2.240}\end{equation}

\noindent and its eigenvalues are 
$-1, -1, -1+{{\vec k}^2\over {\omega}^2(k)}.$

So the metric $\hat\eta$ is proper Riemannian. It can be shown that the 
manifold $H^1_3$ has constant negative curvature and that it is a space-like 
surface, whose normal is 

\begin{equation}
n^{\mu} \equiv ({\omega(k)\over m}, {\vec{k}\over m}) \equiv {k^{\mu}\over m},
\quad n^2 = 1.\label{2.250}\end{equation}

This normal can be extended to a vector field on the Minkowski space $M^4$ 
with

\begin{equation}
n_{q} = {q^{\mu}\over\sqrt{q^2}}{\partial\over \partial q^{\mu}},
\label{2.260}\end{equation}

\noindent with $q^0 > 0$.

The only invariant second order differential operator on $H^1_3$ is the 
Laplace-Beltrami operator $\Delta$ (no invariant differential operator 
of first order exists \cite{Rudin}):

\begin{equation}
\Delta = {1\over \sqrt{\mid\hat\eta\mid}}
{\partial\over \partial k^i}
{\hat\eta}^{ij}\sqrt{{\mid\hat\eta\mid}}
{\partial\over \partial k^j},\label{2.270}
\end{equation}

\noindent where $\mid\hat\eta\mid = det\{\hat\eta_{ij}\}$. 

Explicitly, this operator is 

\begin{equation}
\Delta = -{\omega(k)\over m}{\partial\over \partial k^i}(\delta^{ij} + 
{k^i k^j\over m^2}){m\over \omega(k)}{\partial\over \partial k^j},\label{2.280}
\end{equation}

\noindent or

\begin{equation}
\Delta = -[\vec{\nabla}^2 + {2\over m^2}\vec{k}\cdot\vec{\nabla} + 
{1\over m^2}(\vec{k}\cdot\vec{\nabla})^2].\label{2.290}\end{equation}

This operator is invariant under a Lorentz transformation $\Lambda$ 

\begin{equation}
k^i \rightarrow k^{\prime\ i} = \Lambda^i_j k^j + \Lambda^i_0
\sqrt{m^2+{\vec k}^2},
\label{2.300}\end{equation}

The operator $\Delta$ is formally selfadjoint with respect to the invariant 
measure $\tilde{dk}$. It is an elliptic operator and has the property 

\begin{equation}
\Delta k^{\mu} = - {3\over m^2} k^{\mu}.\label{2.310}\end{equation}

Let us define 

\begin{equation}
D = m^2\Delta + 3.\label{2.320}\end{equation}

\noindent so that 

\begin{equation}
D k^{\mu} = 0.\label{2.330}\end{equation}

When explicitly written (see the following Section) the equation

\begin{equation}
D f(\vec k) = 0,\label{2.340}\end{equation}

\noindent is an equation which can be separated in spherical coordinates. 
The radial equation has three regular singularities in the complex plane 
of $z = -{{\vec k}^2\over m^2}$, which are in $-1, 0, \infty$, so it is a 
hypergeometric equation. The four functions $k^{\mu}$ are a subset of the 
solutions of Eq.(\ref{2.340}), with the characteristic exponents $l = 0,1$ 
in the neighborhood of the point 0 ($k^{\circ}=\omega(k)$ has  the exponent 
$l=0$ and $\vec k$ has the exponent $l=1$). In the neighborhood of the point 
at infinite the characteristic exponents are $+1$ and $-3$. So we may expect 
that an infinite set of solutions could have the same asymptotic 
behaviour like $k^{\mu}$, when $\vert{\vec k}\vert \rightarrow \infty$. 

We will see in the next Section that this is indeed the case and that there 
is an infinite set of solutions of the Eq.(\ref{2.340}) which gives a set of 
well defined integrals, when we put them in place of $k^{\mu}$ in 
Eq.(\ref{2.110}).

\section{The eigenfunctions of the Laplace-Beltrami operator}

The Laplace-Beltrami operator $\Delta$ of Eqs.(\ref{2.270}) and (\ref{2.290}) 
was studied in a series of papers by Raczka, Limic and Niederle \cite{RLN}, 
where is called $\Delta(H^1_3)$ (see Eq.(5.10) in the first reference 
of \cite{RLN}). There it is shown that $\Delta$ has no discrete spectrum, 
but only a continuous one, and the basis of its generalized eigenfunctions 
is determined. 

$\Delta$ can be identified with one of the Casimir operators of the Lorentz 
group, whose Lie algebra is defined in terms of the differential operators

\begin{equation}
l_{\mu\nu} = i D_{\mu\nu},\label{3.100}\end{equation}

\noindent with $D_{\mu\nu}$ as in the Eq.(\ref{2.170}).

To make contact with the notations used by  Naimark in his book \cite{NAI} 
we put

\begin{equation}
L_i = l_i,\quad K_i = - l_{0i},\quad l_i = -{1\over 2}\epsilon_{ijk} l_{jk},
\quad (i,j,k = 1,2,3),\label{3.110}
\end{equation}

\noindent (in Naimark's notations: 

\begin{equation}
H_3 = L_3;\quad H_{\pm} = L_{\pm} = L_1 \pm iL_2;\quad 
F_3 = K_3;\quad F_{\pm} = K_{\pm} = K_1 \pm iK_2.) \label{3.120}
\end{equation}

These operators satisfy the Lorentz algebra

\begin{eqnarray}
[ L_i,\ L_j ] &=& i \epsilon_{ijk} L_k,\nonumber\\  
{[ K_i,\ L_j ]} &=& i \epsilon_{ijk} K_k,\nonumber\\  
{[ K_i,\ K_j ]} &=& - i \epsilon_{ijk} L_k.\label{3.130}
\end{eqnarray}

Two invariant operators can be defined using $\vec{L}$ and $\vec{K}$

\begin{equation}
{\Xi}_1 = \vec{L}\cdot\vec{K},\quad {\Xi}_2 = 
\vert\vec{K}\vert^2 - \vert\vec{L}\vert^2,\label{3.140}
\end{equation}

\noindent and these are the two Casimir operators of Lorentz group. 

The operators ${\vec L}$ and ${\vec K}$ can be written

\begin{equation}
\vec{L} = -i\vec{k}\wedge\vec{\nabla},\quad \vec{K} = i\omega(k) \vec{\nabla},
\label{3.150}
\end{equation}

\noindent so

\begin{equation}
\left\{ \begin{array}{rcl}
\Xi_1 &=& -\left[\vec{k}\wedge\vec{\nabla} \omega(k)\right]\cdot \vec{\nabla},
\\
\Xi_2 &=& m^2{\nabla}^2 + 2\left(\vec{k}\cdot\vec{\nabla}\right) + 
\left(\vec{k}\cdot\vec{\nabla}\right)^2. 
\end{array}\right.
\label{3.160}
\end{equation}

\noindent $\omega(k)$ has a gradient which is parallel to $\vec{k}$, so 
$\Xi_1 = 0$. We are left with one Casimir only 

\begin{equation}
\Xi_2 = \Delta.\label{3.170}
\end{equation}

In terms of $l_{\mu\nu}$ we have

\begin{equation}
\Delta = -{1\over 2} l_{\mu\nu} l^{\mu\nu},\quad\quad 
\epsilon_{\mu\nu\rho\lambda} l^{\mu\nu} l^{\rho\lambda} = 0,\label{3.180}
\end{equation}

\noindent and of course we have

\begin{equation}
[ D,\ D_{\mu\nu}] = 0,\label{3.190}
\end{equation}

\noindent where the operator $D$ is defined in Eq.(\ref{2.320}). 

If we denote $\lambda$ the eigenvalue of $m^2\Delta$, the relation with 
Naimark's notations \cite{NAI} is given by 

\begin{equation}
\lambda = - ( k_{\circ}^2 + c^2 - 1),\quad k_{\circ} c = 0,\label{3.200}
\end{equation}

\noindent where $k_{\circ}$ and $c$ are the eigenvalues of two operators  
$\Delta$ and $\Delta^{\prime}$ defined in the quoted reference  

\begin{eqnarray}
\Delta\vert_{eigenvalue} = -2(k_{\circ}^2 + c^2 - 1)\nonumber\\ 
{\Delta}^{\prime}\vert_{eigenvalue} = -4ik_{\circ} c.\label{3.210}
\end{eqnarray}

If we choose 

\begin{equation}
k_{\circ} = 0,\quad c = i\Lambda,\label{3.220}
\end{equation}

\noindent we get

\begin{equation}
\lambda = 1 + \Lambda^2 \ \in\ [1, +\infty),
\label{3.230}
\end{equation}

\noindent corresponding to the representations of the Lorentz group of the 
principal series, with $k_{\circ} = 0$, which are unitary, 
see \cite{NAI}.

The zero modes of the operator $D$ of Eq.(\ref{2.320}) correspond to 

\begin{equation}
\lambda = - 3.\label{3.240}
\end{equation}

This value of $\lambda$ corresponds to a non unitary representation of 
the Lorentz group, which is reducible but not completely reducible 
as will be shown later. 
In Appendix B we give the details of the determination of this 
representation. Let us show here the expression of the boost $K_3$ and its 
action on the representation $\lambda = -3$. From Eqs.(\ref{3.100}), 
(\ref{3.110}) and (\ref{2.170}) we get 

\begin{eqnarray}
l_{ij} = -\epsilon_{ijk} L_k = i(x^i\partial_j - x^j\partial_i),\nonumber\\
l_{0j} = -K_j = - i\sqrt{1 + \vec{x}^2}\ \partial_j,\label{3.250}
\end{eqnarray}

\noindent where

\begin{equation}
\vec{x} = {\vec{k}\over m}.\label{3.260}
\end{equation}

From these expressions we get 

\begin{equation}
K_3 = i\sqrt{1+ r^2}\left(\cos{\theta}{\partial\over \partial r} - 
{\sin{\theta}\over r}{\partial\over \partial \theta}\right),\label{3.270}
\end{equation}

\noindent in the spherical coordinates $(r, \theta, \phi)$ of $\vec{x}$. 

$D$ becomes 

\begin{equation}
D = m^2\Delta + 3 = \Delta\vert_x + 3,\label{3.280}
\end{equation}

\noindent where

\begin{equation}
\Delta\vert_x = -\left[\left(1+r^2\right){\partial^2\over \partial r^2} + 
({2\over r} + 3r){\partial\over \partial r} - {J^2\over r^2}\right]
\label{3.290}\end{equation}

\noindent and $J^2 = \vec{L}^2$ as usually. 

The eigenfunctions of $\Delta\vert_x$ correspond to 
$\lambda\ \in\ [1,+\infty)$, or $\Lambda \ \in\ [0,+\infty)$ 
as a limit from the upper half-plane of the complex plane of $\Lambda$. 
They are given in \cite{RLN} (also see Eqs.(\ref{B.110}) and (\ref{B.120}) )

\begin{equation}
\begin{array}{rcl}
u_{\lambda, l, m}(r,\theta,\phi) &=& 
N_{\lambda l}\ v_{1,\lambda,l,m}^{(\circ)} = \\
&=& N_{\lambda l}\ r^l \ 
_2 F_1\left({l+1+i\Lambda\over 2}, {l+1-i\Lambda\over 2}; l+{3\over 2}; -r^2
\right) 
Y_{l,m}(\theta,\phi),
\end{array}
\label{3.300}\end{equation}

\noindent where $_2F_1$ is the hypergeometric function, with the three 
parameters of $_2F_1(\alpha,\beta;\gamma; -r^2)$ satisfying the relation 

\begin{equation}
\alpha + \beta + {1\over 2} = \gamma,\label{3.310}
\end{equation}

\noindent and where $l = 0,1,2,...$, and $\mid m\mid \leq l$.

The normalization factor $N_{\lambda l}$ is 

\begin{equation}
N_{\lambda l} = 
{2\pi\over m\sqrt{\Lambda}}\left\vert{\Gamma({l+2-i\Lambda\over 2})
\Gamma({l+1-i\Lambda\over 2})\over {\Gamma(i\Lambda)\Gamma(l+{3\over 2})}}
\right\vert.\label{3.320}
\end{equation}

\noindent for $\Lambda \in [0,\infty)$.

With this normalization $u_{\lambda,l,m}$ is an orthonormal set with respect 
to the scalar product of $L_2(\tilde{dk})$

\begin{equation}
\int\tilde{dk}\overline{u}_{\lambda l m}(r, \theta, \phi)
u_{\lambda^{\prime}, l^{\prime}, m^{\prime}}(r, \theta, \phi) = 
\delta_{l l^{\prime}}\delta_{m m^{\prime}}\delta(\Lambda - \Lambda^{\prime}).
\label{3.330}\end{equation}

The action of the generators $K_3$ and $L_i$ on this basis is

\begin{equation}
K_3 u_{\lambda, l, m} = 
-i\vert l+1+i\Lambda\vert \ C_{l+1,m} u_{\lambda,l+1, m} + 
i\vert l+i\Lambda\vert \ C_{l,m} u_{\lambda,l-1, m},\label{3.340}
\end{equation}

\noindent where the last term in the r.h.s. is zero when $l=0$ and 
$\mid m\mid$ is always bounded by $l$ and, in the last term, by $l-1$; 

\begin{equation}
C_{l, m} = \sqrt{(l+m)(l-m)\over (2l-1)(2l+1)},\label{3.350}\end{equation}

\noindent and

\begin{equation}
\begin{array}{rcl}
L_3 u_{\lambda,l,m} &=& m u_{\lambda,l,m},\nonumber\\
L_{\pm}u_{\lambda,l, m} &=& \sqrt{l(l+1)-m(m\pm 1)}\ u_{\lambda,l, m\pm 1},
\label{3.360}\end{array}
\end{equation}

\noindent with

\begin{equation}
L_{\pm} = L_1 \pm i L_2.\label{3.370}\end{equation}

The action of $K_1$, $K_2$ can be recovered from the commutators

\begin{equation}
K_{\pm} = K_1 \pm i K_2 = \pm [ K_3,\ L_{\pm} ].\label{3.380}
\end{equation}

Observe that, if we define 

\begin{equation}
K_3 u_{\lambda,l,m} = A_{\lambda, l, m} u_{\lambda,l+1, m} + 
B_{\lambda, l, m} u_{\lambda,l-1, m},\label{3.390}
\end{equation}

\noindent we have

\begin{equation}
B_{\lambda, l, m} = {\overline A}_{\lambda, l-1, m},\label{3.400}
\end{equation} 

\noindent when $\Lambda$ is real.This is the hermiticity condition for 
the $K_3$ matrix on the $u_{\lambda,l,m}$ basis when 
$\lambda\ \in\ [1,+\infty)$.

This basis is complete in $L_2(\tilde{dk})$, that is

\begin{equation}
\sum_{l\geq 0}\sum_{\mid m\mid\leq l}\int_1^{\infty} d\lambda 
u_{\lambda, l, m}(r,\theta,\phi) 
\overline{u}_{\lambda, l, m}(r^{\prime},\theta^{\prime},\phi^{\prime}) = 
\Omega(k)\delta^3(k - k^{\prime}).\label{3.410}
\end{equation}

The expressions given up to now, that is Eqs.(\ref{3.300}), (\ref{3.320}), 
(\ref{3.330}), (\ref{3.340}), (\ref{3.400}) and Eq.(\ref{3.410}) 
hold for $\Lambda\ \in\ [0, +\infty)$, or $\lambda\ \in\ [1, +\infty)$, 
which corresponds to the continuous spectrum of $\Delta$. Since we are 
interested in the case $\lambda = -3$, we need the relations analogous 
to Eq.(\ref{3.340}) for a generic complex $\Lambda$, with 
${\rm Im}\Lambda \geq 0$. 

For a complex $\Lambda$, and in particular for an immaginary $\Lambda$, 
the expression of the normalization factor $N_{\lambda l}$ 
becomes meaningless. Moreover, as discussed in the Appendix B, 
for $\lambda = - 3$ the functions $u_{\lambda,l,m}$ are no more 
normalizable. 

Following the discussion of Appendix B the only set of solutions of 
Eq.(\ref{2.340}) which we will take into account in the following is 
the set $\{ v_{1,\lambda,l,m}^{(\circ)}\}$, with $\lambda = - 3$. 

The action of $K_3$ on $v^{(\circ)}_{1,\lambda,l,m}$ is 

\begin{equation}
K_3 v^{(\circ)}_{\lambda,l,m} = - i{(l+1+i\Lambda)(l+1-i\Lambda)\over 2l+3} 
C_{l+1,m} v^{\circ}_{\lambda,l+1,m} + i(2l+1) C_{l,m} 
v^{\circ}_{\lambda,l-1,m},\label{3.420}
\end{equation}

\noindent as given in Eq.(\ref{B.340}). 

The action of $K_{\pm}$ is given in Eq.(\ref{B.350}), where the dependence 
on $\lambda$ of the coefficients in the r.h.s. is determined by the action 
of $K_3$ of Eq.(\ref{3.420}). 

From this equation we see that, for complex values of $\Lambda$ in the 
upper half-plane, the first term in the r.h.s. of Eq.(\ref{3.420}) 
vanishes for 

\begin{equation}
\Lambda = i (l+1).\label{3.430}
\end{equation}

We may conclude that the {\underbar{only representation}} with an invariant 
subspace of dimension {\underline{4}} is the representation with 

\begin{equation}
\Lambda = 2 i,\quad {\rm or}\quad \lambda = - 3.\label{3.440}
\end{equation}

In the next Section we will study in more detail this representation.

\section{The representation $\lambda = - 3$}

In this Section we will study in some detail the representation 
$\lambda = -3$. For $\Lambda = 2 i$ let us define

\begin{equation}
w_{l,m}(r,\theta,\phi) = v^{\circ}_{-3,l,m}(r,\theta,\phi).\label{4.100}
\end{equation}

From Eq.(\ref{3.420}) we have the action of $K_3$ on $w_{l,m}$: 

\begin{equation}
K_3 w_{l,m} = -i {(l-1)(l+3)\over (2l+3)}\ C_{l+1,m}\ w_{l+1,m} + 
i(2l+1)\ C_{l,m}\ w_{l-1,m},\label{4.110}
\end{equation}

\noindent where the coefficients $C_{l,m}$ are defined in Eq.(\ref{3.350}), 
and where the values of $\mid m\mid$ in the right hand side 
are always limited by $l$ in the first term and by $l-1$ in the second term. 

Explicitly $w_{\l,m}(\vec x)$ is given by 

\begin{equation}
w_{l,m}(r,\theta,\phi) = 
r^l\ _2F_1\left({l-1\over 2},{l+3\over 2};l+{3\over 2}; -r^2\right)\ 
Y_{l,m}(\theta,\phi),\label{4.120}
\end{equation}

\noindent where $r, \theta$ and $\phi$ are the spherical coordinates of 
$\vec x = {{\vec k}\over m}$. 

Observe that for $l=0,1$ 

\begin{eqnarray}
w_{0,0} = \sqrt{1+r^2}\ Y_{0,0} = {\omega(k)\over m}\ Y_{0,0},\nonumber\\
w_{1,m} = r\ Y_{1,m} = {\mid\vec{k}\mid\over m}\ Y_{1,m},\label{4.130}
\end{eqnarray}

\noindent which are the components of the 4-vector ${k^{\mu}\over m}$ in 
the spherical basis, with the Minkowski diagonal metric given by 

\begin{equation}
\tilde\eta = {4\pi m^2\over 3}\ (3, -1, -1, -1),\label{4.140}
\end{equation}

\noindent where $(l,m) = {(0,0),(1,-1),(1,0),(1,-1)}$ and 

\begin{equation}
\sum_{l=0,1}\sum_{\mid m\mid\leq l} \overline{w}_{l,m} \tilde\eta_{lm,lm} 
w_{l,m} \equiv \overline{w}\cdot\tilde\eta\cdot w 
= m^2.\label{4.150}
\end{equation}

It is important to observe that all the functions $w_{l,m}$ have the 
same asymptotic behaviour as $\mid\vec{k}\mid \rightarrow \infty$, and not 
only those for $l=0,1$.  In other words we have an infinite sequence of 
zero modes of the operator $D$, all with the same asymptotic behaviour 
as $k^{\mu}$. This fact is explicitly shown by Eq.(\ref{B.211}).

The subspace spanned by the component $l=0,1$ of $w_{l,m}$ is 
{\sl invariant} under the action of the generators of the Lorentz 
group. Indeed, from the Eq.(\ref{4.110}), we have

\begin{eqnarray}
K_3 w_{0,0} &=& {i\over \sqrt{3}} w_{1,0},\nonumber\\
K_3 w_{1,m} &=& i \sqrt{3}\ \delta_{m,0} w_{0,0}.\label{4.160}
\end{eqnarray}

Moreover, since 

\begin{equation}
K_{\pm} = \pm [ K_3,\ L_{\pm}],\label{4.170}
\end{equation}

\noindent and $L_{\pm}$ doesn't modify the value of $l$, even the action 
of $K_{\pm}$ leaves the subspace $l=0,1$ invariant.

It is the factor $l-1$ in the r.h.s. of Eq.(\ref{4.110}) which is responsible 
of this fact. There are no other values of $\Lambda$, in the upper half of its 
complex  plane, which could provide an invariant subspace of the same 
dimension.  If we choose $\Lambda = in$, with $n$ integer $> 0$, 
we have an invariant subspace with a dimension which is determined by the 
maximum value of $l$, which is equal to $n-1$, as we can see from 
Eq.(\ref{3.420}). In conclusion, the value $\lambda = - 3$, or 
$\Lambda = 2 i$, is determined by the requirement of the existence of an 
invariant subspace of dimension 4.  

The representation whose basis is $\{w_{l,m}\}$ is  reducible, but not 
decomposable (not completely reducible). That it is reducible is clear, since 
the matrix of $K_3$, determined by Eq.(\ref{3.420}), is block triangular. 
Indeed, if we define 

\begin{equation}
K_3 w_{l,m} = w_{l',m} (K_3)_{l',l},\label{4.180}
\end{equation}

\noindent we have 

\begin{equation}
(K_3)_{2,1} = 0,\label{4.190}
\end{equation}

\noindent but 

\begin{equation}
(K_3)_{1,2} \not= 0.\label{4.191}
\end{equation}
 
Now, a representation like this is decomposable if there exists a 
similarity transformation defined by a matrix S of the form 

\begin{equation}
\left(\matrix{1&Y\cr 0&1\cr}\right),\label{4.200}
\end{equation}

\noindent transforming all the 
generators of the Lorentz group $l_{\mu\nu}$ in a block diagonal form, 
see \cite{NS}. 

The matrix $Y$ has four rows and infinitely many columns and of course it must 
be non-zero. It should similarly transform in a block diagonal form also 
the matrix of ${\vec L}^2$, which is already in block diagonal form. This 
is clearly impossible. So the representation is indecomposable.

\section{The supertranslations}

The presence of an entire series of zero modes of the operator $D$, with 
the same asymptotic behaviour as $\mid\vec k\mid \rightarrow \infty$, 
allows us to define the following set of integrals 

\begin{equation}
P_{l,m} = \int{\tilde{dk}} w_{l,m}(\vec k) {\overline{a}}(\vec k) a(\vec k),
\label{5.100}
\end{equation}

\noindent where 

\begin{equation}
{\overline{P}}_{l,m} = (-1)^m P_{l,-m}, 
\label{5.101}
\end{equation}

\noindent and where the functions $w_{l,m}$ are given in 
Eq.(\ref{4.120}).

As shown in Appendix B (see Eq.(\ref{B.211})) all these integrals are well 
defined. 

The canonical action of the generators of the Lorentz group $M^{\mu\nu}$ 
on the $P_{l,m}$ can be obtained from Eqs.(\ref{2.150}) and (\ref{2.160}); 
since from Eq.(\ref{2.170}) we have

\begin{equation}
\left\{ \begin{array}{rcl}
D^{ij} &=& ( k^i{\partial\over \partial k^j} - 
k^j{\partial\over \partial k^i}) = 
i\epsilon_{ijk} L_k,\\  
D^{0j} &=& \omega(k){\partial\over \partial k^j} = -i K_j,\\ 
\end{array}\right. \label{5.110}
\end{equation}

\noindent with the definitions (\ref{3.100}) and (\ref{3.110}), we get 

\begin{equation}
\{ M^{i,j},\ P_{l,m}\} = -i\epsilon_{ijk}P_{l^{\prime},m^{\prime}} 
(L_k)_{l^{\prime},m^{\prime};l,m},\label{5.120}
\end{equation}

\begin{equation}
\{ M^{0,j},\ P_{l,m}\} = i P_{l^{\prime},m^{\prime}} 
(K_j)_{l^{\prime},m^{\prime};l,m}.\label{5.121}
\end{equation}

In these equations the matrices $\parallel L_k\parallel$ and 
$\parallel K_j\parallel$ are those given in Eqs.(\ref{B.270}), (\ref{B.390}) 
and (\ref{B.400}), and correspond to the representation $\lambda = -3$ of 
the Lorentz algebra. 

Moreover we have

\begin{equation}
\{ P_{l,m},\ P_{l',m'}\} = 0.\label{5.130}
\end{equation}

For $l=0,1$ the $P_{l,m}$ are the spherical components of $P^{\mu}$. 

The Eqs.(\ref{5.120}), (\ref{5.121}) and (\ref{5.130}) show that this 
algebra is infinite-dimensional, that the abelian subalgebra of the 
translations and supertranslations $\{ P_{l,m}\}$ is a normal subalgebra, 
and that the factor algebra is isomorphic to the Lorentz algebra. 

In order to show that this algebra is the same as that of the BMS group, 
we simply redefine the $P_{l,m}$ with 

\begin{equation}
{\hat P}_{l,m} = \nu_l P_{l,m},\label{5.140}
\end{equation}

\noindent with $\nu_l$ satisfying the recurrence relation 

\begin{equation}
\nu_{l+1} = {2l+3\over l+3} \nu_l,\label{5.150}
\end{equation}

\noindent or 

\begin{equation}
\nu_l = {(2l+1)!!\over (l+2)!} 2\nu_0.\label{5.160}
\end{equation}

By defining 

\begin{equation}
R_z = i K_3,\quad R_{\pm} = i K_{\pm},\quad L'_z = i L_z,\label{5.170}
\end{equation}

\noindent we get exactly the algebra given by Sachs \cite{SACHS} 
(see in this reference the Eqs.(IV.19) and (IV.20)). 

As shown by Sachs \cite{SACHS}, the 4-dimensional subgroup of translations 
($P_{l,m}$ with $l=0,1$) is unique. On the other hand the homogeneous 
Lorentz group is not similarly unique. This is due to the fact that copies 
of the Lorentz group can be obtained by a conjugation 
with an arbitrary supertranslation. 

Indeed, let us define the following transformation of the Lorentz 
generators

\begin{eqnarray}
M^{\prime i,j} \rightarrow M^{\prime i,j} + 
\alpha \{ P_{l,m},\ M^{\prime i,j}\} = \nonumber\\ 
= M^{\prime i,j} + i \alpha \epsilon_{ijk}P_{l^{\prime},m^{\prime}}
(L_k)_{l^{\prime},m^{\prime};l,m},  \label{5.180}
\end{eqnarray}

\begin{eqnarray}
M^{\prime 0,j} \rightarrow M^{\prime 0,j} + 
\alpha \{ P_{l,m},\ M^{\prime 0,j}\} = \nonumber\\ 
= M^{\prime 0,j} - i \alpha P_{l^{\prime},m^{\prime}}
(L_j)_{l^{\prime},m^{\prime};l,m}, 
\label{5.190}
\end{eqnarray}

\noindent where $\alpha$ is an arbitrary real parameter, and 
Eqs.(\ref{5.120}), (\ref{5.121}) were used.

For a fixed supertranslation $P_{l,m}$ we may exponentiate the infinitesimal 
transformation (\ref{5.180}), (\ref{5.190}), getting the same r.h.s. terms: 

\begin{equation}
e^{\alpha P_{l,m}}* M^{\prime i,j} = M^{\prime i,j} +
i\alpha \epsilon_{ijk}P_{l^{\prime},m^{\prime}}
(L_k)_{l^{\prime},m^{\prime};l,m},\label{5.191}
\end{equation}

\begin{equation}
e^{\alpha P_{l,m}}*M^{\prime 0,j} = M^{\prime 0,j} -
i \alpha P_{l^{\prime},m^{\prime}}
(L_j)_{l^{\prime},m^{\prime};l,m}.\label{5.192}
\end{equation}

In these equations the $*$ operation is defined by 

\begin{equation}
e^A *B = \sum_{n\geq 0}{1\over n!} D_A^n B,\quad D_A = \{ A,\ \cdot\}.
\label{5.193}
\end{equation}

This transformation corresponds to a conjugation of the Lorentz 
algebra with an arbitrary fixed supertranslation. It can be verified that 
the transformed algebra is again the Lorentz algebra. 

As a consequence, the Casimir operator of the Poincar\'e group given by the 
square of the Pauli-Lubanski four-vector, which is invariant under the 
transformation (\ref{5.191}) and (\ref{5.192}) when $l=0,1$, will in general 
change \cite{SACHS}. 

The canonical action of a fixed supertranslation on the field is determined 
by  

\begin{equation}
\{ P_{l,m},\ a(\vec k)\} = i w_{l,m}(\vec k) a(\vec k),\quad 
\{ P_{l,m},\ {\overline{a}}(\vec k)\} = 
-i w_{l,m}(\vec k) {\overline{a}}(\vec k),\label{5.200}
\end{equation}

\noindent which, for $l=0,1$, reduces to (\ref{2.150}), written in spherical 
coordinates. 

Since $P_{l,m}$ is not real ($\overline{P}_{l,m} = (-1)^m P_{l,-m}$), it 
induces two different canonical transformations on the field $\Phi$, 
determined by its real and its imaginary part 

\begin{equation}
P_{l,m} = R_{l,m} + i I_{l,m},\label{5.201} 
\end{equation}

\noindent where 

\begin{equation}
\{ R_{l,m},\ a(\vec k)\} = i Re(w_{l,m}(\vec k))\ a(\vec k),\quad 
\{ R_{l,m},\ {\overline{a}}(\vec k)\} = -i Re(w_{l,m}(\vec k))\  
{\overline{a}}(\vec k),\label{5.202}
\end{equation}

\noindent and 

\begin{eqnarray}
\{ I_{l,m},\ a(\vec k)\} = i Im(w_{l,m}(\vec k))\  a(\vec k),\quad  
\{ I_{l,m},\ {\overline{a}}(\vec k)\} = -i Im(w_{l,m}(\vec k))\ 
{\overline{a}}(\vec k).\label{5.203}
\end{eqnarray}

Again these can be exponentiated for a fixed supertranslation, for instance 

\begin{eqnarray}
e^{\alpha R_{l,m}} a(\vec k) = e^{i\alpha Re w_{l,m}(\vec k)} a(\vec k),\quad 
e^{\alpha R_{l,m}} {\overline{a}}(\vec k) = 
e^{- i\alpha Re w_{l,m}(\vec k)} {\overline{a}}(\vec k),\label{5.210}
\end{eqnarray}

\noindent and the analogous expressions for $I_{l,m}$. 

The Eq.(\ref{5.210}) defines a canonical transformation of the field, since 
the canonical Poisson brackets (\ref{A.260}) are invariant under such 
transformation. 

This transformation is non-local on the field $\Phi({\vec x},t)$. It is a 
particular case of a linear transformation 

\begin{equation}
(\Phi,\Pi) \rightarrow (\Phi',\Pi'),\label{5.240}
\end{equation}

\noindent realized as an integral transformation of convolution type. Indeed, 
$a(\vec k)$ belongs to $L^2(R^3)$, and so is a tempered distribution, 
and the exponentials ${\rm exp}(i Re w_{l,m})$ and ${\rm exp}(i Im w_{l,m})$ 
are tempered distributions too. The convolution product between their 
Fourier transforms is well defined. Indeed, the functions $w_{l,m}(\vec k)$, 
being the solutions of a homogeneous elliptic equation, are infinitely 
differentiable \cite{Rudin}. They and all their derivatives are polynomially 
bounded, due to the bound (\ref{B.211}). So, the functions $w_{l,m}(\vec k)$ 
and their exponentials ${\rm exp}(i Re w_{l,m})$, ${\rm exp}(i Im w_{l,m})$, 
are multipliers in $S^{\prime}(R^3)$, the space of tempered distributions on 
$R^3$ \cite{Gel1}. This implies that their Fourier transforms are convolutes 
\cite{Gel2}, that is it exists their convolution with any tempered 
distribution. In conclusion  we may write the relation among $(\Phi,\Pi)$ 
and $(\Phi',\Pi')$ as 

\begin{eqnarray}
\Phi^{\prime}(\vec x, t) = 
\int d^3x^{\prime} [ f(\vec x - \vec x^{\prime})\Phi(\vec x, t) 
+ g(\vec x - \vec x^{\prime}) \Pi(\vec x^{\prime}, t)],\nonumber\\ 
\Pi^{\prime}(\vec x, t) = 
\int d^3x^{\prime} [ h(\vec x - \vec x^{\prime})\Phi(\vec x, t) 
+ k(\vec x - \vec x^{\prime}) \Pi(\vec x^{\prime}, t)],
\label{5.250}
\end{eqnarray}

\noindent where the distributions $f$, $g$ and $h$ are defined by   

\begin{eqnarray}
f(\vec x) &=& \int \tilde{dk} \omega(k) \left( e^{i\alpha Re w_{l,m}(\vec k) + 
i{\vec k}\cdot{\vec x}} + c.c\right),\nonumber\\ 
g(\vec x) &=& i \int \tilde{dk} \left( e^{i\alpha Re w_{l,m}(\vec k) + 
i{\vec k}\cdot{\vec x}} - c.c\right),\nonumber\\ 
h(\vec x) &=& -i \int \tilde{dk} {\omega}^2(k) 
\left( e^{i\alpha Re w_{l,m}(\vec k) + i{\vec k}\cdot{\vec x}} - c.c\right), 
\nonumber\\ 
k(\vec x) &=& f(\vec x).\label{5.260}
\end{eqnarray}

\noindent where c.c means the complex conjugated. It is easily seen that 
the condition 

\begin{equation}
\int d^3y \left[ f(\vec x - \vec y) k({\vec x}^{\prime} -  \vec y) - 
g(\vec x - \vec y) h({\vec x}^{\prime} - \vec y)\right] = 
\delta^3(\vec x- \vec x^{\prime}),\label{5.261}
\end{equation}

\noindent which must be satisfied if the transformation (\ref{5.250}) has to 
be canonical, holds. 

The transformation induced by $I_{l,m}$ is the same with the replacement 
$Re w_{l,m} \rightarrow Im w_{l,m}$. 

The transformation (\ref{5.250}) changes the initial configuration of the 
field determined by the functions $a(\vec k)$, to a new one, determined by 
the functions $a^{\prime}(\vec k) = {\rm exp}(i\alpha Re(w_{l,m}(\vec k))) 
a(\vec k)$. As we have seen this transformation can change the spin content 
of the field, when $l\geq 2$.

As a conclusion we collect the Eqs.(\ref{5.120}), (\ref{5.121}) and 
(\ref{5.130}), which give the Poisson algebra of the BMS group

\begin{eqnarray}
\{ M^{\prime \mu\nu},\ P_{l,m}\} &=& P_{l^\prime, m^{\prime}} 
(M^{\prime \mu\nu})_{l^{\prime},m^{\prime};l,m},\nonumber\\ 
\{ P_{l,m},\ P_{l^{\prime},m^{\prime}}\} &=& 0,\label{5.270}
\end{eqnarray}

\noindent where, as given by Eqs(\ref{5.120}) and (\ref{5.121}) 

\begin{eqnarray}
(M^{\prime ij})_{l^{\prime},m^{\prime};l,m} &=& -i\epsilon_{ijk} 
(L_k)_{l^{\prime},m^{\prime};l,m},\nonumber\\ 
(M^{\prime 0j})_{l^{\prime},m^{\prime};l,m} &=& i 
(K_j)_{l^{\prime},m^{\prime};l,m},\label{5.280}
\end{eqnarray}

\noindent with the matrix elements of $L_k$ and of $K_j$ given in 
Eqs.(\ref{B.270}), (\ref{B.390}) and (\ref{B.400}). 

\section{Conclusions}

Following our initial purpose, we have found a Poisson algebra isomorphic 
to the BMS algebra, realized on the phase space of a real classical 
Klein-Gordon field. 

The structure constants of this algebra are given by the matrix elements 
of an infinite-dimensional representation of the Lorentz group, which 
is non unitary, reducible and indecomposable. We have given an explicit 
basis of functions for this representation, which are not normalizable 
in the sense of an $L_2$ space with Lorentz invariant measure, but for 
which we give an asymptotic bound. 

The requirement of the existence of a four-dimensional invariant subspace 
selects this representation almost uniquely among all the possible 
representations of the Lorentz group. 

Thus, we have shown that it is possible to realize the BMS algebra outside 
the general relativity context, in which it was originally 
discovered. This fact suggests a more general role of the BMS group. 

As it is well known the BMS group is the semidirect product of the Lorentz 
group with the finite set of translations and the infinite set of 
supertranslations. In the general relativity case, on the basis of 
physical arguments, the vanishing of the supertranslations is required 
\cite{SACHS} \cite{BMSREVIEW}. 
In the case of the same algebra realized in terms of the Klein-Gordon 
field this requirement should be a strong restriction on the possible field 
configurations, which could not have a clear justification. 

As we have observed in the Introduction, we may expect that a similar 
analysis could be worked out for other classical massive fields. Something 
similar should also happen for a zero mass field, by performing a harmonic 
analysis on the $k_{\mu} k^{\mu} = 0$ cone, where the Laplace-Beltrami 
operator changes significantly \cite{RLN1}. 

The existence of the supertranslations in the context of the Klein-Gordon 
field theory is a byproduct of a larger study on the search of a canonical 
set of collective and relative variables for a classical relativistic 
field, playng a role analogous to the center of mass and relative variables 
of a non relativistic system of particles. This study will be the argument 
of a forthcoming paper.

\acknowledgments

The authors wish to thank L. Lusanna for many suggestions and enlightening 
discussions on the subject of this paper, for his useful criticism and for 
reading the manuscript. The authors are also indebted with K. Marathe for 
many suggestions and for reading the manuscript.

\appendix
\section{Notations for the classical real Klein-Gordon field}

  We list in this appendix the various definitions concerning the real 
Klein-Gordon field. We will put $c= \hbar = 1$, and the metric 
signature is $(+;-,-,-)$. 

The Lagrangian and the lagrangian density are

\begin{equation}
L = \int dt {\cal L},\quad\quad 
{\cal L} = {1\over 2}({\partial_{\mu}\Phi\partial^{\mu}\Phi} - m^2\Phi^2),
\label{A.100}
\end{equation}

\noindent where $\mu = 0,1,2,3$ and 

\begin{equation}
{\overline{\Phi}} = \Phi.\label{A.110}
\end{equation}

\noindent The conjugate momentum and the equation of motion are

\begin{equation}
\Pi(x) ={\dot\Phi}(x) \equiv {\partial_{\circ}\Phi(x)},\quad\quad 
(\Box + m^2)\Phi(x) = 0.\label{A.120}\end{equation}

The N\"oether currents associated to a Poincar\'e transformation are given 
by

\begin{eqnarray}
j^{\mu\nu} &=& \partial^{\mu}\Phi\partial^{\nu}\Phi - 
\eta^{\mu\nu}{\cal L} = j^{\nu\mu},\label{A.130}\\
j^{\mu 0} &=& {\dot\Phi}\partial^{\mu}\Phi = \Pi\partial^{\mu}\Phi,\quad 
(\mu \neq 0),\label{A.140}\\
j^{00} &=& {1\over 2}[{\dot \Phi}^2 + (\vec{\nabla}\Phi)^2 + m^2\Phi^2],
\label{A.150}
\end{eqnarray}

\noindent and the generators of the Poincar\'e group are

\begin{equation}
P^{\mu} = \int d^3x j^{0\mu}(x),\quad\quad M^{\mu\nu} = 
\int d^3x (x^{\mu} j^{0\nu} - x^{\nu} j^{0\mu}).\label{A.160}
\end{equation}

The canonical Poisson brackets are (see reference \cite{HAM})

\begin{equation}
\{\Phi(\vec{x}, x^{0}),\ \Pi(\vec{x}^{\prime},x^0)\} = 
\delta^3(\vec{x} - \vec{x}^{\prime}),\label{A.170}
\end{equation}

\noindent with the other Poisson brackets vanishing.

The Poincar\'e algebra is

\begin{equation}
\{P^{\mu},\ P^{\nu}\} = 0,\label{A.180}
\end{equation}

\begin{equation}
\{M^{\mu\nu},\ P^{\rho}\} = P^{\mu}\eta^{\nu\rho} - P^{\nu}\eta^{\mu\rho},
\label{A.190}
\end{equation}

\begin{eqnarray}
\{M^{\mu\nu},\ M^{\rho\lambda}\} = 
M^{\mu\lambda}\eta^{\nu\rho} + 
\eta^{\mu\lambda}M^{\nu\rho}-\nonumber\\
- M^{\mu\rho}\eta^{\nu\lambda}-
\eta^{\mu\rho}M^{\nu\lambda}.\label{A.200}
\end{eqnarray}

The Fourier expansion of the field is

\begin{equation}
\Phi(\vec{x},t) = \int\tilde{dk}[a(\vec{k}) e^{-i(k\cdot x)} + 
{\overline{a}}(\vec{k}) e^{i(k\cdot x)}],\label{A.210}
\end{equation}

\noindent where c.c. means the complex conjugate, and $(\ \cdot \ )$ is 
the usual Lorentz invariant scalar product between 4-vectors, and 
(we use the notations of \cite{IZ})

\begin{eqnarray}
\tilde{dk} &=& {d^3 k\over \Omega(k)},\quad \Omega(k) = 
(2\pi)^3 2 \omega(\vec{k}),\nonumber\\
{\vec k} &\equiv& \{ k^i\},\quad (i=1,2,3),\quad k_i =- k^i,\nonumber\\ 
\quad\omega(k) &=& k_{\circ} = 
\sqrt{\vec{k}^2 + m^2}.\label{A.220}
\end{eqnarray}

If we denote with $\hat\Phi(\vec k, t)$ the Fourier transform of the field, 
with respect to the measure $d^3 k$, then 

\begin{equation}
\Phi(\vec k, t) = \int d^3k {\hat{\Phi}}(\vec k, t) e^{i{\vec k}\cdot{\vec x}},
\quad {\rm with} \quad 
{\overline{\hat\Phi}}(\vec k, t) = {\hat\Phi}(-\vec k, t),\nonumber
\end{equation}

\noindent and 

\begin{equation}
a(\vec k) = {1\over 2} \Omega(k) e^{i\omega(k) t} 
\left[ {\hat{\Phi}}(\vec k, t) + 
{i\over \omega(k)} {\dot{\hat{\Phi}}}(\vec k, t)\right].\label{A.230}
\end{equation}

\noindent From this we have the bound:

\begin{equation}
\vert a(\vec k)\vert^2 \leq (2\pi)^6\left[ 
{\vec k}^2\vert{\hat{\Phi}}(\vec k, t)\vert^2 + 
\vert{\dot{\hat{\Phi}}}(\vec k, t)\vert^2\right].\label{A.240}
\end{equation}

Since we have assumed that $\Phi(\vec x, \cdot)$, $\dot{\Phi}(\vec x, \cdot)$ 
and ${\vec\nabla}\Phi(\vec x, \cdot)$ be functions in $L^2(R^3)$, from 
the Parseval identity we get that $a(\vec k) \in L^2(R^3)$. 

This implies the existence of $P^{\mu}$, but not of $M^{\mu\nu}$. This 
last it will be assured if we assume ${\vec\nabla} a(\vec k) \in L^2(R^3)$. 
So $a(\vec k)$ will go to zero as $\mid\vec k\mid\rightarrow \infty$, as in 
Eq.(\ref{2.120}). 

The conjugate momentum is 

\begin{equation}
\Pi(x) = -i\int\tilde{dk}\omega(k)[a(\vec{k})e^{-i(k\cdot x)} - 
{\overline{a}}(\vec{k})e^{i(k\cdot x)}].
\label{A.250}\end{equation}

The Poisson brackets for the Fourier coefficients are 

\begin{equation}
\{a(\vec{k}), \overline{a}(\vec{k}^{\prime})\} = 
-i\Omega(k)\delta^3(\vec{k} - \vec{k}^{\prime}),\label{A.260}
\end{equation}

\noindent with the other Poisson brackets vanishing. 

The Fourier coefficients in terms of the field are given by 

\begin{equation}
a(\vec{k}) = \int d^3x e^{i(k\cdot x)}[\omega(\vec{k})\Phi(x) + i \Pi(x)].
\label{A.270}
\end{equation}

In terms of the Fourier coefficients the Poincar\'e generators are the 
following:

\begin{equation}
P^{\mu} = \int\tilde{dk} k^{\mu}\overline{a}(\vec{k}) a(\vec{k}),
\label{A.280}
\end{equation}

\begin{equation}
M^{ij} = -i\int\tilde{dk}\overline{a}(\vec{k})(k^i{\partial\over 
\partial k^j} - 
k^j{\partial\over \partial k^i}) a(\vec{k}) = {\overline{M}}^{ij},
\label{A.290}
\end{equation}

\begin{equation}
M_{0j} = 
t P_j - i\int\tilde{dk} \overline{a}(\vec{k})\omega(\vec{k})
{\partial\over \partial k^j} a(\vec{k}) = - M^{jo} = {\overline{M}}^{0j}.
\label{A.300}
\end{equation}

We define

\begin{eqnarray}
M^{\prime 0j} &=& -i\int{\tilde{dk}} {\overline{a}}(\vec k)\omega(k){\partial
\over \partial k^j} a(\vec k),\nonumber\\
M^{\prime ij} &=& M^{ij},\label{A.310}
\end{eqnarray}

\noindent where $M^{\prime ij}$ and $M^{\prime 0j}$ has the same Poisson 
algebra as $M^{ij}$ and $M^{0j}$, and will be used as the Lorentz generators.

\section{The eigenfunctions of the Laplace-Beltrami operator and the 
action of the Lorentz generators}

\subsection{The eigenfunctions of the Laplace-Beltrami operator}

The eigenfunctions of the continuous spectrum of the Laplace-Beltrami 
operator, corresponding to $\lambda\ \in\ [1, \infty)$, have been
studied in \cite{RLN}. However, we are interested even in the 
non-normalizable solutions of the equation

\begin{equation}
\left( \Delta - \lambda\right) v_{\lambda,l,m} = 0,\label{B.100}
\end{equation}

\noindent for different values of $\lambda$.

A fundamental system of solutions, in the neighborhood of the origin, that is 
for $r\simeq 0$,  of the Eq.(\ref{B.100}) in spherical coordinates, is

\begin{eqnarray}
v^{(\circ)}_{1,\lambda,l,m}(\vec{r}) = u_{1,\lambda,l}^{(\circ)}(r) 
Y_{l,m}(\theta,\phi),\nonumber\\
v^{(\circ)}_{2,\lambda,l,m}(\vec{r}) = u_{2,\lambda,l}^{(\circ)}(r) 
Y_{l,m}(\theta,\phi),\label{B.110}
\end{eqnarray}

\noindent where $r = {\mid\vec{k}\mid\over m}$, $Y_{l,m}$ are the 
spherical harmonics as defined in \cite{MES}, and 

\begin{eqnarray}
u_{1,\lambda,l}^{(\circ)}(r) = 
r^l\ _2 F_1({l+1+i\Lambda\over 2},{l+1-i\Lambda\over 2};l+{3\over 2}; -r^2),
\nonumber\\
u_{2,\lambda,l}^{(\circ)}(r) = 
r^{-l-1}\ _2 F_1(-{l+i\Lambda\over 2},-{l-i\Lambda\over 2};{1\over 2}-l; -r^2),
\label{B.120}\end{eqnarray}

\begin{equation}
(u^{(\circ)}_{2,\lambda,l}(r) = u^{(\circ)}_{1,\lambda,-l-1}(r)),\label{B.130}
\end{equation}

The relation between $\lambda$ and $\Lambda$ is given by 
$\lambda = 1 + \Lambda^2$, with $Im \Lambda \geq 0$, as  in Section III. 

A fundamental system in the neighborhood of the point at infinite 
$r\simeq\infty$ is 

\begin{eqnarray}
v^{(\infty)}_{1,\lambda,l,m}(\vec{r}) = u_{1,\lambda,l}^{(\infty)}(r) 
Y_{l,m}(\theta,\phi),\nonumber\\
v^{(\infty)}_{2,\lambda,l,m}(\vec{r}) = u_{2,\lambda,l}^{(\infty)}(r) 
Y_{l,m}(\theta,\phi),\label{B.140}
\end{eqnarray}

\noindent where

\begin{eqnarray}
u_{1,\lambda,l}^{(\infty)}(r) = r^{-1-i\Lambda}\  
_2 F_1({l+1+i\Lambda\over 2},-{l-i\Lambda\over 2};1+i\Lambda; -{1\over r^2}),
\nonumber\\
u_{2,\lambda,l}^{(\infty)}(r) = r^{-1+i\Lambda}\  
_2 F_1({l+1-i\Lambda\over 2},-{l+i\Lambda\over 2};1-i\Lambda; -{1\over r^2}).
\label{B.150}\end{eqnarray}

For $r\geq 0$ no other singular points are met. For the argument 
$z = -1$, the hypergeometric series is absolutely convergent, since 
the coefficients of $_2 F_1(\alpha,\beta;\gamma; z)$ satisfy the condition 
(see Eq.(\ref{3.310}))

\begin{equation}
\alpha + \beta - \gamma = -{1\over 2}.\label{B.160}
\end{equation}

If $i\Lambda$ is a positive integer, the solution 
$u_{1,\lambda,l}^{(\infty)}(r)$ requires a modification of the expression 
given in the Eq.(\ref{B.150}). 
However, since we are only interested in the solution 
$u_{1,\lambda,l}^{(\circ)}(r)$ for ${\rm Im}\Lambda \geq 0$, we do not 
give here the necessary modification. 

We are interested in the normalization properties of these solutions in the 
neighborhood of the points $0$ and $\infty$, with respect to the invariant 
measure

\begin{equation}
\tilde{dr} = {d^3 r\over \sqrt{1+r^2}}.\label{B.170}
\end{equation}

The solution  $v^{(\circ)}_{1,\lambda,l,m}$ is regular and normalizable in 
the neighborhood of the origin. The solution $v^{(\circ)}_{2,\lambda,l,m}$ 
on the other hand is normalizable in the neighborhood of 
origin for $l=0,1$ only. We will discard this solution, even in the case 
$l=0,1$, since under the action of the Lorentz generators it will be 
transformed into a solution with a different value of $l$, thus becoming 
non-normalizable in the neighborhood of origin. See also the discussion 
at the end of subsection 2.

The solution $u^{(\circ)}_{1,\lambda,l}$ can be analytically continued to 
$r\simeq\infty$, and, for a generic value of $\Lambda$ has an asymptotic 
expansion which is a linear combination of the two power of $r$ (see 
Eq. 2.10(2) of \cite{ERD})

\begin{equation}
r^{-1-i\Lambda},\quad r^{-1+i\Lambda},\label{B.180}
\end{equation}

\noindent and, for a real $\Lambda$, is normalizable in the sense of the 
continuous spectrum. 

For real values of $\Lambda \in[0,\infty)$, $v^{(\circ)}_{1,\lambda,l,m}$  
is an eigenfunction of the operator $\Delta$, see \cite{RLN}, 
belonging to the continuous spectrum. 

In the case $\lambda = - 3$, that is $\Lambda = 2 i$, we must use another 
asymptotic expansion. In general, for $\Lambda = n i$, with n integer, 
we must use the expansion given in Eq. 2.10(7) of \cite{ERD}; for 
$\Lambda = 2 i$ and for $r\rightarrow\infty$ we get 

\begin{equation}
u_{1,-3,l}^{(\circ)} \simeq r + O(r^{-1}).\label{B.190}
\end{equation}

So, this solution is no more normalizable at $\infty$, but only at $0$.

The other solution at $r\simeq \infty$ has the behaviour

\begin{equation}
u_{2,-3,l}^{(\circ)} \simeq r + O(r^{-1}),\quad {\rm for}\quad l=0,1;\quad  
u_{2,-3,l}^{(\circ)} \simeq r + O(r^{-3}),\quad {\rm for}\quad l\geq 2.\quad  
\label{B.200}\end{equation}

\noindent and is singular at the origin for $l>0$. So it is not normalizable 
at $r \simeq \infty$.

As for the second fundamental system given in Eq.(\ref{B.140}) 
and (\ref{B.150}) they are linear combinations of the first set, since 
they can be obtained by analytic continuation using well known relations. 
The solution $u^{(\infty)}_{1,\lambda,l}$ has, for $r\simeq\infty$, 
the asymptotic behaviour 

\begin{equation}
r^{-1+{\rm Im}\Lambda-i{\rm Re}\Lambda},\label{B.210}
\end{equation}

\noindent and, for ${\rm Im}\Lambda \geq 0$, is not normalizable. 

The second solution $u^{(\infty)}_{2,\lambda,l,m}$ is instead normalizable 
for $r \simeq \infty$. 

The solutions $w_{l,m} = v^{(\circ)}_{1,-3,l,m}$ satisfy the following 
inequality

\begin{equation}
\vert w_{l,m}(\vec k)\vert \leq \sqrt{{2l+1\over 4\pi}} \sqrt{1+r^2} M_l,
\label{B.211}
\end{equation}

\noindent where

\begin{eqnarray}
M_l &=& {\Gamma(l+{3\over2}) \Gamma(2)\over 
\Gamma({l\over 2}+2) \Gamma({l\over 2})},\quad {\rm if}\quad l\geq 2,\nonumber\\ 
M_l &=& 1,\quad {\rm if}\quad l=0,1.\label{B.212}
\end{eqnarray}

Indeed, since

\begin{equation}
\vert Y_{l,m}(\theta,\phi)\vert \leq \sqrt{{2l+1\over 4\pi}},\label{B.213}
\end{equation}

\noindent and

\begin{eqnarray}
u_l(r) &=& r^l \ _2F_1({l-1\over 2}, {l+3\over 2}; l+{3\over2}; -r^2) = 
\nonumber\\
&=& r^l (1+r^2)^{{1-l\over 2}} \ _2F_1({l-1\over 2}, {l\over 2}; l+ {3\over 2};
{r^2\over 1+r^2}),\label{B.214}
\end{eqnarray}

\noindent using Eq.2.9(3) of \cite{ERD}, and the bounds

\begin{equation}
0 \leq {r^2\over 1+r^2} < 1,\label{B.215}
\end{equation}

\begin{equation}
\vert _2F_1(\alpha, \beta; \gamma; x)\vert \leq F(\alpha, \beta; \gamma; 1) = 
{\Gamma(\gamma - \alpha - \beta) \Gamma(\gamma)\over 
\Gamma(\gamma - \alpha) \Gamma(\gamma - \beta)},\label{B.216}
\end{equation}

\noindent which hold if $\alpha , \beta, \gamma >0$ and 
$\gamma - \alpha - \beta > 0$, we have from Eq.(\ref{B.214}) 

\begin{equation}
\vert u_l(r)\vert \leq r^l (1+r^2)^{{1-l\over 2}} 
{\Gamma(2)\Gamma(l+{3\over 2})\over \Gamma({l\over 2}+2)\Gamma({l+3\over 2})}.
\label{B.217}
\end{equation}

Now

\begin{equation}
r^l (1+r^2)^{{1-l\over 2}} \leq \sqrt{1+r^2},\label{B.230}
\end{equation}

\noindent and collecting the results we get the inequality (\ref{B.211}) 
for $l\geq 2$.

For $l = 0,1$ the last hypergeometric function in Eq.(\ref{B.214})
is equal 1, and we get the inequality (\ref{B.211}) for $l=0,1$.

The inequality (\ref{B.211}) for $w_{l,m}$ implies the existence of the 
integrals $P_{l,m}$ defined in Eq.(\ref{5.100}). Indeed

\begin{equation}
\vert P_{l,m}\vert \leq \int{\tilde{dk}} \vert w_{l,m}(\vec k)\vert 
{\overline{a}}(\vec k) a(\vec k) \leq \sqrt{{2l+1\over 4\pi}} M_l 
{P_{\circ}\over m}.\label{B.240}
\end{equation}

So, if $P_{\circ}$ exists, all the integrals $P_{l,m}$ will similarly exist.

\subsection{The action of the Lorentz generators}

Let us determine the action of the generators of the Lorentz group on 
the solutions $v^{(\circ)}$. 

The explicit expression of the generators is

\begin{eqnarray}
L_3 &=& -i{\partial\over \partial \phi},\nonumber\\
L_{\pm} &=& L_1 \pm iL_2 = e^{\pm i\phi}\left(\pm{\partial\over \partial \theta} 
+i \cot{\theta}{\partial\over \partial \phi}\right),\nonumber\\
K_3 &=& i\sqrt{1+r^2}\left(\cos{\theta}{\partial\over \partial r} - 
{\sin{\theta}\over r}{\partial\over \partial\theta}\right),\nonumber\\
K_{\pm} &=& i\sqrt{1+r^2} e^{\pm i\phi}
\left(\sin{\theta}{\partial\over \partial r} + 
{\cos{\theta}\over r}{\partial\over \partial \theta} \pm 
{i\over r\sin{\theta}}{\partial\over \partial \phi}\right).\label{B.250}
\end{eqnarray}

They satisfy the algebra

\begin{eqnarray}
{\left[ L_+,\ L_-\right]} &=& 2 L_3,\quad 
{\left[ L_3,\ L_{\pm}\right]} = \pm L_{\pm},\nonumber\\
{\left[ K_+,\ K_-\right]} &=& -2 L_3,\quad 
{\left[ K_3,\ K_{\pm}\right]} = {\mp} L_{\pm},\nonumber\\
{\left[ K_3,\ L_3\right]} &=& 0,\quad 
{\left[ K_3,\ L_{\pm}\right]} = {\pm} K_{\pm},\nonumber\\
{\left[ L_3,\ K_{\pm}\right]} &=& {\pm} K_{\pm},\quad 
{\left[ K_{\pm},\ L_{\pm}\right]} = 0.\label{B.260}
\end{eqnarray}

The action of $L_3$ and $L_{\pm}$ is the usual one

\begin{equation}
\begin{array}{rcl} 
L_3 v_{..,\lambda,l,m} &=& m v_{..,\lambda,l,m},\\
L_{\pm} v_{..,\lambda,l,m} &=& \sqrt{l(l+1)-m(m\pm 1)} v_{..,\lambda,l,m\pm 1}.
\label{B.270}
\end{array}\end{equation}

For the action of $K_3$ and $K_{\pm}$ we will use the following formulas

\begin{equation}
{d\over dr}F(a_l,b_l;c_l;-r^2) = -{r\over 2\sqrt{1+r^2}}{(2a_l)(2b_l)\over c_l}
F(a_{l+1},b_{l+1};c_{l+1};-r^2),\label{B.280}
\end{equation}

\begin{eqnarray}
&{d\over dr}\left[r^l F(a_l,b_l;c_l;-r^2)\right] = \nonumber\\
&=(2l+1){r^{l-1}\over \sqrt{1+r^2}}F(a_{l-1},b_{l-1};c_{l-1};-r^2) - \nonumber\\
&- (l+1) r^{l-1} F(a_l,b_l;c_l;-r^2),\label{B.290}
\end{eqnarray}

\noindent where $F(a_l,b_l;c_l;-r^2)$ is the hypergeometric function and 
where, see Eq.(\ref{3.330}) 

\begin{equation}
a_l + b_l +{1\over 2} = c_l.\label{B.300}
\end{equation}

The Eq.(\ref{B.280}) can be obtained by using the Eqs.2.11(10) and 2.8(20) of 
\cite{ERD}, and the Eq.(\ref{B.290}) using the Eqs.2.8(27) and 2.11(10) 
of the same reference.

Moreover, we will use the properties of the spherical harmonics \cite{MES} 

\begin{equation}
\cos{\theta} Y_{l,m}(\theta,\phi) = C_{l+1,m} Y_{l+1,m}(\theta,\phi) + 
C_{l,m} Y_{l-1,m}(\theta,\phi),\label{B.310}
\end{equation}

\noindent where 

\begin{equation}
C_{l,m} = \sqrt{(l-m)(l+m)\over (2l-1)(2l+1)},\label{B.320}
\end{equation}

\noindent and

\begin{eqnarray}
\sin{\theta}{\partial\over \partial\theta} Y_{l,m}(\theta,\phi) = 
l\cos{\theta} Y_{l,m}(\theta,\phi) - \nonumber\\
- \sqrt{{(2l+1)(l^2-m^2)\over (2l-1)}}
Y_{l-1,m}(\theta,\phi).\label{B.330}
\end{eqnarray}

With these relations we get

\begin{eqnarray}
K_3 v^{(\circ)}_{1,\lambda,l,m} =& 
-i{(l+1+i\Lambda)(l+1-i\Lambda)\over (2l+3)} C_{l+1,m} 
v^{(\circ)}_{1,\lambda,l+1,m}+\nonumber\\ 
&+i(2l+1) C_{l,m} v^{(\circ)}_{1,\lambda,l-1,m},\label{B.340}
\end{eqnarray}

\noindent and 

\begin{eqnarray}
K_{\pm} v^{(\circ)}_{1,\lambda,l,m} =& 
{\pm} i [ {(l+1+i\Lambda)(l+1-i\Lambda)\over 
(2l+3)}\sqrt{{(l+1\pm m)(l+2\pm m)\over (2l+1)(2l+3)}} 
v^{(\circ)}_{1,\lambda,l+1,m\pm 1}+ \nonumber\\
&+(2l+1)\sqrt{{(l\mp m)(l\mp m-1)\over (2l-1)(2l+1)}}
v^{(\circ)}_{1,\lambda,l-1,m\pm 1}] .\label{B.350}
\end{eqnarray}

For the solution $v_{2,\lambda,l,m}$ we get similarly

\begin{eqnarray}
K_3 v^{(\circ)}_{2,\lambda,l,m} =&
+i{(l+i\Lambda)(l-i\Lambda)\over (2l-1)} C_{l,m} 
v^{(\circ)}_{2,\lambda,l-1,m} - \nonumber\\
&-i(2l+1) C_{l+1,m} v^{(\circ)}_{2,\lambda,l+1,m},\label{B.360}
\end{eqnarray}

\noindent and

\begin{eqnarray}
K_{\pm} v^{(\circ)}_{2,\lambda,l,m} =& 
{\pm} i [ {(l+i\Lambda)(l-i\Lambda)\over (2l-1)}
\sqrt{{(l{\mp}m)(l-1{\mp}m)\over (2l-1)(2l+1)}} 
v^{(\circ)}_{2,\lambda,l-1,m{\pm}1}+ \nonumber\\
&+(2l+1)\sqrt{{(l+1{\pm}m)(l+2{\pm}m)\over (2l+1)(2l+3)}}
v^{(\circ)}_{2,\lambda,l+1,m{\pm}1}] .\label{B.370}
\end{eqnarray}

If we put $\Lambda = 2i$ we get for 

\begin{equation}
w_{l,m} = v^{(\circ)}_{1,-3,l,m},\quad\quad w^{\prime}_{l,m} = 
v^{(\circ)}_{2,-3,l,m},\label{B.380}
\end{equation}

\begin{eqnarray}
K_3 w_{l,m} =& 
-i{(l-1)(l+3)\over (2l+3)} C_{l+1,m} w_{l+1,m} + \nonumber\\
&+i(2l+1) C_{l,m} w_{l-1,m},\label{B.390}
\end{eqnarray}

\begin{eqnarray}
K_{\pm} w_{l,m} =& {\pm} i [ {(l-1)(l+3)\over 
(2l+3)}\sqrt{{(l+1\pm m)(l+2\pm m)\over (2l+1)(2l+3)}} w_{l+1,m\pm 1}
+ \nonumber\\
&+(2l+1)\sqrt{{(l\mp m)(l\mp m-1)\over (2l-1)(2l+1)}}w_{l-1,m\pm 1}] .
\label{B.400}
\end{eqnarray}

\begin{eqnarray}
K_3 w^{\prime}_{l,m} =& 
+i{(l-2)(l+2)\over (2l-1)} C_{l,m} w^{\prime}_{l-1,m} - \nonumber\\
&-i(2l+1) C_{l+1,m} w^{\prime}_{l+1,m},\label{B.410}
\end{eqnarray}

\begin{eqnarray}
K_{\pm} w_{l,m}^{\prime} =& {\pm} i [ {(l-2)(l+2)\over (2l-1)}
\sqrt{{(l{\mp}m)(l-1{\mp}m)\over (2l-1)(2l+1)}} w_{l-1,m{\pm}1}^{\prime}+ 
\nonumber\\
&+(2l+1)\sqrt{{(l+1{\pm}m)(l+2{\pm}m)\over (2l+1)(2l+3)}}
w_{l+1,m{\pm}1}^{\prime}] .\label{B.420}
\end{eqnarray}

In the previous formulas, the terms on the right hand side with the function 
$w_{l,m}$ with $l$ negative or with $\mid m\mid > l$ are to be considered zero.

Referring to the Eq.(\ref{B.390}) and (\ref{B.400}), observe that the 
representation subspace corresponding to the values of $l = 0,1$ is an 
invariant subspace. This is due to the factor $(l-1)$. 

Similarly, looking at the Eq.(\ref{B.410}) and (\ref{B.420}), we see that 
the subspace corresponding to the values $l\geq 2$ is invariant. 

Let us define the matrices corresponding to the representations $w_{l,m}$ 
and $w^{\prime}_{l,m}$ 

\begin{equation}
M w_{l,m} = w_{l^{\prime},m^{\prime}} (M)_{l^{\prime},m^{\prime};l,m},\quad 
M w^{\prime}_{l,m} = 
w^{\prime}_{l^{\prime},m^{\prime}} 
(M^{\prime})_{l^{\prime},m^{\prime};l,m},\label{B.430}
\end{equation}

\noindent where $M$ is any one of the Lorentz generators. 

If we define the new bases

\begin{equation}
\hat{w}_{l,m} = N_l w_{l,m},\quad {\rm and} \quad 
\hat{w}^{\prime}_{l,m} = N^{\prime}_l w^{\prime}_{l,m},\label{B.440}
\end{equation}

\noindent with 

\begin{equation}
N_l = N^{\prime}_l = {1\over \sqrt{2l+1}},\label{B.450}
\end{equation}

\noindent we get for the new matrices 

\begin{equation}
(\hat{M})_{l^{\prime},m^{\prime};l,m} = 
{N_l\over N_{l^{\prime}}} (M)_{l^{\prime},m^{\prime};l,m},\quad {\rm and}\quad 
(\hat{M^{\prime}})_{l^{\prime},m^{\prime};l,m} = 
{N_l\over N_{l^{\prime}}} 
(M^{\prime})_{l^{\prime},m^{\prime};l,m},\label{B.460}
\end{equation}

\noindent the relation

\begin{equation}
(\hat{M}^{\prime})^{\dag} = \hat{M}.\label{B.470}
\end{equation}

That is the representation $\hat{w}^{\prime}_{l,m}$ is the adjoint of 
the representation $\hat{w}_{l,m}$ \cite{NAI}

\end{document}